\input psfig 
\documentstyle[twocolumn,prl,floats,aps]{revtex}

\draft

\begin{document}

\twocolumn[\hsize\textwidth\columnwidth\hsize\csname @twocolumnfalse\endcsname

\title{\bf Microscopic Oscillations in the Quantum Nucleation of Vortices \\
Subject to Periodic Pinning Potential in a Thin Superconductor }

\author{ Roberto Iengo \\ {\it International School for Advanced Studies, Via
Beirut 4, 34014 Trieste (Italy) \\ and INFN -- Sezione di Trieste, 34100
Trieste (Italy) } }

\author{ Giancarlo Jug \\ {\it INFM and  Istituto di Scienze Matematiche,
Fisiche e Chimiche \\ Universit\`a di Milano a Como, Via Lucini 3, 22100 Como
(Italy) \\ and INFN -- Sezione di Pavia, 27100 Pavia (Italy) } }

\date{ 23 October 1995 }

\maketitle

\begin{abstract}

We present a theory for the decay of a supercurrent through nucleation of
vortex-antivortex pairs in a two-dimensional superconductor in the presence of
dissipation and of a periodic pinning potential. Through a powerful quantum
electrodynamics formulation of the problem we show that the nucleation rate
develops oscillations in its current-density dependence which are connected to
the pinning periodicity. A remnant of the dissipation-driven localization
transition is present, and an estimate of the nucleation rate suggests that
these effects might be observable in real thin superconductors.

\end{abstract}


]

There has been a strong revival of interest, recently, in the physics of
magnetic vortices in type II and high-temperature superconductors
\cite{reviews}. Phenomena involving such vortices pose well-defined problems
which are of great importance in the technological applications of the new
high-$T_c$ materials. Of particular interest are the statics (the vortex-glass
transition) and dynamics (flux creep, vortex tunneling) of vortex systems in
the presence of pinning impurity centers and an externally applied
supercurrent. Most of these issues are connected with the residual resistence
due to the thermal or quantum motion of the vortices and as hindered by the
presence of pinning centers.

In a recent article,\cite{ieju} we have pointed out that such residual
resistence should develop even in the absence of an applied magnetic field.
This is due to the spontaneous nucleation in the bulk of vortex-antivortex
pairs induced, in the presence of a supercurrent driven by an external source,
by the electromagnetic-like Magnus force acting on the quantum-mechanical
vortices thought of as electron-positron-like pairs. This quantum nucleation
occurs homogeneously throughout the sample. Exploiting this analogy with
quantum electrodynamics (QED), a powerful ``relativistic'' quantum field theory
(QFT) approach has been set up to study vortex nucleation in the
two-dimensional (2-D) geometry and in the presence of quantum dissipation and
of infinite barrier-height pinning centers.  The central result is that the
nucleation rate $\Gamma$ has a strong exponential dependence on the number
current density $J$, given by $\Gamma{\propto}\eta^{1/2}\eta_{eff}J^{-1}
\exp\{-\eta_{eff}{\cal E}_{0R}^2/4\pi J^2\}$. Here, $\eta_{eff}$ is an
effective viscosity coefficient as renormalised by the magnetic-like part of
the Magnus force, and ${\cal E}_{0R}$ is the rest or nucleation energy of a
single vortex as renormalized by screened Coulomb interactions and (fake)
Landau-level corrections. This exponential dependence would make the vortex
nucleation (folded, e.g., into the sample's resistance) observable in a rather
narrow range of $J$-values (see below).  We stress that whilst the possibility
of homogeneous vortex nucleation as a decay mechanism for a supercurrent was
independently and first proposed by Ping Ao, \cite{ao} the extension to
dissipative dynamics, leading to qualitatively different results, was first
treated in our recent paper, \cite{ieju} to which the interested reader is
referred for details of the formalism.

In this Letter we extend our formulation to the important situation in which
finite barrier-height pinning centers are present and we initially envisage a
geometry in which the pinning potential is periodic, with a lattice spacing
$d=2\pi/k$ and finite amplitude $U_0$. We estimate the order of magnitue of the
production rate $\Gamma$ to find that under reasonable conditions the
nucleation phenomenon ought to be observable.  Moreover, we show that a remnant
of the dissipation-induced vortex localization transition \cite{schmid,ghm} is
present through a $J$- dependence of $\Gamma$ in which microscopic (or,
depending on $d$, mesoscopic) oscillations are noted, due to the underlying
periodicity of the pinning potential, and might also be observable under
suitable conditions. Beside spontaneous bulk nucleation in the absence of a
magnetic field, our formalism and results can also be applied to vortex
tunneling from the edge, extending the work of Ref.
\cite{aoth93,aoth94,stephen} with the aim of obtaining an explicit formula,
complete of prefactors, for the tunneling rate. Details of our calculations for
the periodic pinning potential and the treatment for the edge tunneling problem
will be presented elsewhere.

We begin by summarising the problem at hand, essentially a single- particle
one, through the classical equation of motion for the vortex moving in the
$(x-y)$ plane of a 2-D superconducting film with a current flowing in the
$y$-direction:

\begin{equation} m\ddot{\bf q}=-{\nabla}U({\bf q})+e{\bf E}\cdot{\bf
q}-e\dot{\bf q}{\times} {\bf B}-\eta\dot{\bf q} \label{classic} \end{equation}

\noindent Here, $m{\rightarrow}0$ is the inertial mass of the vortex of
topological charge $e=\pm1$ and 2-D coordinate ${\bf q}(t)$ subject to the
Magnus force having an electric-like ${\bf E}=2\pi J\hat{\bf x}$ as well as a
magnetic-like ${\bf B}=2\pi^2\rho_s s\hat{\bf z}$ ($\rho_s$ being the
superfluid 3-D number density and $s$ the film thickness). The e.m. current
density $J^{em}$ flowing through the sample's cross section is related to the
number current density via $J=J^{em} s/2e$. We work in the units system in
which $\hbar =1$.  Also, $\eta$ is the friction coefficient and $U({\bf q})$
the phenomenological periodic pinning potential which for convenience we take
to be of the cosine form with wavenumber ${\bf Q}=(k,k)$: \cite{note1}

\begin{equation} U({\bf q}(t))=U_0 \sum_{a=1}^D \left [ 1 - \cos \left (
kq_a(t) \right ) \right ] \label{potent} \end{equation}

\noindent The quantum-mechanical version of Eq. (\ref{classic}) is attained
through the Feynman path-integral (FPI) transposition in which dissipation is
treated quantistically through the formulation due to Caldeira and Leggett.
\cite{cale} Particle nucleation (closely related to tunneling) is dealt with
through a scalar-QED, 2-D version of Schwinger's calculation for the vacuum
decay rate $\Gamma$ in the presence of the crossed uniform fields $({\bf
E},{\bf B})$. The formula for $\Gamma$ reads \cite{ieju}

\begin{equation} \frac{\Gamma}{L^D}=\frac{2}{L^DT} Im \int_{\epsilon}^{\infty}
\frac{d\tau}{\tau} e^{-{\cal E}_0^2\tau} \int {\cal D}q(t) \exp\{
-\int_0^{\tau} dt {\cal L}_E \} \label{rate} \end{equation}

\noindent where $L^DT$, with $D=2$ here, is the space-time volume of the
sample, ${\cal E}_0$ is the vortex nucleation energy, and the Euclidean
single-particle Lagrangian for closed trajectories in space-time is

\begin{eqnarray} {\cal
L}_E&=&\frac{1}{2}m_{\mu}\dot{q}_{\mu}\dot{q}_{\mu}-\frac{1}{2}i
\dot{q}_{\mu}F_{\mu\nu}q_{\nu} + V({\bf q}) \nonumber \\ &+&\sum_k \left \{
\frac{1}{2}m_k\dot{\bf x}_k^2 +\frac{1}{2}m_k\omega_k^2 \left( {\bf
x}_k+\frac{c_k}{m_k\omega_k^2}{\bf q} \right )^2 \right \} \label{lagran}
\end{eqnarray}

\noindent Here, the set of $\{ {\bf x}_k \}$ represents the Caldeira-Leggett
bath of harmonic oscillators simulating quantum friction, with
$\frac{\pi}{2}\sum_k\frac{c_k^2}{m_k\omega_k}\delta (\omega-\omega_k)
=\eta\omega\exp(-\omega/\Omega)$ ($\Omega$ being some frequency cutoff) fixing
the spectral distribution of frequencies (we restrict the present discussion to
the ohmic case). Also, $F_{\mu\nu}$ is the uniform Magnus field tensor and
$V({\bf q})=2{\cal E}_0U({\bf q})$ is the relativistic QFT counterpart of the
classical potential, Eq. (\ref{potent}), thus having amplitude $A_0=2{\cal
E}_0U_0$ (we have assumed $A_0/{\cal E}_0^2{\ll}1$).  Also,
$m_3=\slantfrac{1}{2}$ and $m_1=m_2=\slantfrac{\gamma}{2}$ with
$\slantfrac{1}{\gamma}={\cal E}_0/m{\rightarrow}\infty$ playing the role of the
square of the speed of light in this formalism, where the non-relativistic
limit corresponds to the overdamped case (which we consider).

Before delving into the evaluation of the FPI in Eq. (\ref{rate}), we briefly
dwell on the physics of the problem at hand in the absence of the Magnus force,
and on the meaning of Eq. (\ref{rate}).  As is well-known, \cite{schmid,ghm}
the motion of a quantum particle in a periodic potential in the presence of
friction displays a localization transition whereby for $\eta < \eta_c$ the
particle is mobile whilst for $\eta > \eta_c$ it is confined,
$\eta_c=k^2/2\pi=2\pi/d^2$ being a sharp threshold independent of the
potential's amplitude $A_0$. This is reproduced by our formalism, with ${\bf
E}={\bf B}=0$ and $\gamma{\rightarrow}0$ in Eq. (\ref{lagran}), with the
confined phase treated within the variational ``blocked-renormalization''
scheme first introduced by Fisher and Zwerger, \cite{fizw} through which a
``mass'' is dynamically generated for the damped modes' propagator. When the
Magnus force-field is turned on (at least its electric-like part) a pole
develops in the bare propagator, signalling the onset of mobility for all
values of $\eta$. This divergence is in turn embedded in the formula for the
pair production rate, Eq. (\ref{rate}). This one comes from a formulation of
vortex nucleation \cite{ieju} in terms of a scalar quantum field $\phi$ with
QFT Lagrangian

\begin{equation} {\cal L}(\phi)=D_0\phi^{*}D_0\phi - \frac{1}{\gamma} {\bf
D}\phi^{*}\cdot{\bf D}\phi - \left ( {\cal E}_0^2 + V({\bf q}) \right )
\phi^{*}\phi \label{qft} \end{equation}

\noindent where $D_{\mu}=\partial_{\mu}-ieA_{\mu}$ with
$A_{\mu}=\frac{1}{2}F_{\mu\nu}q_{\nu}$, in which the process of
particle-antiparticle pair production is quite naturally encoded. In this QFT
formulation, the vacuum decay rate is computed through the FPI on all closed
loops in space-time weighted with the single-particle Euclidean Lagrangian
corresponding to (\ref{qft}).

We now come to the evaluation of $\Gamma$, in the presence of the periodic
potential which calls for a renormalization group (RG) type approach. Indeed,
apart from the subtle question of dynamic mass generation for $\eta > \eta_c$,
the problem at hand is that of a sine-Gordon-like model in (1+0) dimensions.
Integrating out the Euclidean ``time''-like component $q_3(t)$, we reach a
formulation in which the electric-like and the magnetic-like Magnus field
components are disentangled. In terms of Fourier components $\bar{\bf
q}_n=\bar{\bf q}(\omega_n)$, with $\omega_n=2\pi n/\tau$:

\begin{eqnarray} &&\int_0^{\tau} dt {\cal
L}_E(q_{\mu}){\rightarrow}\int_0^{\tau} dt {\cal L}_E({\bf q})=\tau \sum_{n\neq
0} \{ \frac{1}{4}\gamma\omega_n^2+\frac{1}{2}\eta |\omega_n| \nonumber \\ && -
E^2{\delta}_{a1} \} \bar{q}_a(\omega_n) \bar{q}_a(-\omega_n) +\int_0^{\tau} dt
V({\bf q}) + {\rm magn.~~part} \label{lagranr} \end{eqnarray}

\noindent In this paper, we take the stance that the only role of the
magnetic-like field in the presence of dissipation is to renormalize the
nucleation energy, ${\cal E}_0^2{\rightarrow}{\cal E}_{0R}^2={\cal
E}_0^2+\frac{1}{\gamma} \sqrt{\eta^2+B^2}$, as well as the friction
coefficient, $\eta{\rightarrow}\eta_{eff}=(\eta^2+B^2)/\eta$ (denoted by $\eta$
in what follows). This was indeed one of the main results of our investigation,
\cite{ieju} and it amounts to an effective reduction of the $D$-dimensional
problem to a one-dimensional system in the presence of the sole electric-like
field. \cite{stephen} A crucial role is played by dissipation, of course, as
otherwise we would be confronted with the Azbel-Hofstadter problem in the
presence of both periodic pinning and magnetic field. The evaluation of the FPI
proceeds by means of the integration over the zero-mode, $\bar{q}_0$, and the
high-frequency modes, $\bar{q}_n$ with $n>1$, where we take the point of view
that, as for the free particle in the uniform field, \cite{ieju} the leading
term in $\Gamma$ comes from the divergence of the FPI associated with the
lowest mode coupling to ${\bf E}$. We apply this scheme to the formula for the
rate, Eq. (\ref{rate}), where the effective 1-D Lagrangian is the same as Eq.
(\ref{lagranr}), with $\gamma =0$ and renormalized ${\cal E}_0$ and $\eta$, and
taking $q_1{\equiv}q$ to be strictly a one-dimensional coordinate.

The effect of the modes $\bar{q}_n$ with $n > 1$ is taken into account through
a frequency-shell RG-like argument by which, for the interesting $\eta >
\eta_c$ case, these modes simply renormalize $A_0$ and are integrated out as
effectively ``massive'' \cite{fizw}, that is via a Lagrangian ${\cal
L}(\bar{q}_n)=\tau \left ( \frac{1}{2}\eta |\omega_n| +M^2-E^2 \right )
\bar{q}_n\bar{q}_n^{*}$ (we will see below that at the relevant singularity
$E^2=\frac{1}{2}\eta|\omega_1|$, so that effectively $n{\rightarrow}n-1$). The
result of this integration is an additional, entropy-like renormalization of
the activation energy, to be included in ${\cal E}_{0R}$, beside the
renormalization of $A_0$. Notice that in the mobile phase, $\eta < \eta_c$, the
pinning potential's amplitude would renormalize to zero, suppressing the
oscillations (to be discussed below) in the vortex nucleation rate. We are
therefore left with the integration over the modes $\bar{q}_0$ and $\bar{q}_1$,
with a potential term

\begin{eqnarray} &&-\int_0^{\tau} dt V_R(q_0,q_1(t)) =A_{0R}\int_0^{\tau} dt
\cos k(q_0+q_1(t)) - A_0\tau \nonumber \\ &&{\simeq}A_{0R}\tau
J_0(2k|\bar{q}_1|)\cos kq_0 - A_0\tau \label{potentr} \end{eqnarray}

\noindent In this equation, $J_0$ is the Bessel function and the new amplitude
$A_{0R}$, as renormalised by the ``blocked'' $n > 1$ modes in the confined
$\eta > \eta_c$ phase, is given by

\begin{eqnarray} A_{0R}= \left \{ \begin{array}{ll} A_0 \left (
\frac{\Omega\tau}{2\pi} \right )^{-\eta_c/\eta} & \mbox{if $\eta < \eta_c$} \\
A_0 \left ( \frac{\mu +1}{ \mu +n^{*-1} } \right )^{-\eta_c/\eta} & \mbox{if
$\eta > \eta_c$} \end{array} \right.  \label{amplitr} \end{eqnarray}

\noindent where we take the cutoff integer
$n^{*}(\tau)=[\Omega\tau/2\pi]{\gg}1$ and a dimensionless mass
$\mu=\frac{2M^2}{\Omega\eta}=\left ( \frac{2\pi A_0 \eta_c}{\Omega\eta} \right
)^{\eta/\eta-\eta_c}$ has been introduced. Notice that, from the variational
approach, $\mu=0$ for $\eta <\eta_c$. Also, in Eq. (\ref{potentr}) the phase of
the $\bar{q}_1$ mode has been integrated out (a procedure generating an
expansion controlled by the non-divergent amplitude $A_{0R}$), allowing us to
integrate out the $\bar{q}_0$-mode exactly. This leads to the expression

\begin{eqnarray} &&\frac{\Gamma}{2L^2}= Im \int_{\epsilon}^{\infty} d{\tau}
{\cal N}(\tau) e^{-({\cal E}_{0R}^2+A_0)\tau} \times \nonumber \\ &&
\int_0^{\infty} d|\bar{q}_1|^2 e^{-(\pi\eta-E^2\tau)|\bar{q}_1|^2} I_0 \left (
A_{0R}\tau J_0(2k|\bar{q}_1|) \right ) \label{rate1} \end{eqnarray}

\noindent where ${\cal N}(\tau)$ is a suitable normalization factor and $I_0$
is the modified Bessel function. The vortex nucleation rate, therefore, bears
some sign of the friction-driven localization transition, through $A_{0R}\tau$
as given by Eq. (\ref{amplitr}). It is clear that the singularity in $\tau$
from the $\bar{q}_1$-integral occurs at $\tau_1=\pi\eta/E^2$; taking a suitable
$i\epsilon$ prescription \cite{ieju} we get

\begin{equation} \frac{\Gamma}{2L^2}=\frac{ {\cal N}(\tau_1) }{ 4\pi J^2 }
e^{-({\cal E}_{0R}^2+A_0)\eta /4\pi J^2} I_0 \left ( \frac{A_{0R}\eta}{4\pi
J^2} J_0(2k{\ell}_N) \right ) \label{ratef} \end{equation}

\noindent where we have set $E=2\pi J$. The presence of the $J_0(2k{\ell}_N)$
argument in the correction factor due to the pinning lattice thus gives rise to
oscillations in $\Gamma$ (hence in the sample's resistence) through the
parameter $2k{\ell}_N=4\pi{\ell}_N/d$. Vortex nucleation (or tunneling) is
therefore sensitive to the corrugation of the pinning substrate.

In order to compute the nucleation length ${\ell}_N$ we have used a
saddle-point procedure in Eq. (\ref{rate1}), noticing that for large $\tau$:
$I_0(x\tau){\rightarrow}\frac{1}{\sqrt{2\pi |x|\tau}}e^{|x|\tau}$. This leads
to a self-consistent equation, ${\cal E}_{0R}^2+A_0 - A_{0R}|J_0(2k{\ell}_N)| =
4\pi^2 J^2 {\ell}_N^2$, which we solve to first order, neglecting corrections
for small $J$:

\begin{equation} {\ell}_N^2{\simeq}\frac{ {\cal E}_{0R}^2+A_0 }{4\pi^2 J^2}
-\frac{A_{0R}}{4\pi^2 J^2} \left | J_0 \left ( k \frac{ \sqrt{{\cal
E}_{0R}^2+A_0} } {\pi J} \right ) \right | \label{nuclen} \end{equation}

As for the normalization factor ${\cal N}(\tau_1)$, we get ($e$ is here Euler's
number)

\begin{equation} {\cal N}(\tau_1)=\frac{e^2}{16\pi^2}n^{*}(\tau_1)(1+\mu)(1+\mu
n^{*}(\tau_1)) (\pi/\tau_1)^{3/2}\eta^2 \label{normaliz} \end{equation}

\noindent Our final expression for the rate $\Gamma$ (with ${\ell}_N$ given by
Eq.(\ref{nuclen})) is

\begin{eqnarray} &&\Gamma=\Gamma_0K(J)  \nonumber \\
&&\frac{\Gamma_0}{L^2}=\frac{ e\Omega\eta^{3/2} }{ 32\pi^2J } e^{ - \left (
{\cal E}_{0R}^2+A_0 \right ) \eta/4\pi J^2 } \\ \label{final} &&K(J)=e(1+\mu)
\left ( 1+\frac{\mu\Omega\eta}{8\pi^2 J^2} \right ) I_0 \left (
\frac{A_{0R}\eta}{4\pi J^2} J_0(2k{\ell}_N) \right ) \nonumber \end{eqnarray}

\noindent where $K(J)$ is the oscillating correction factor due to the pinning
substrate.

\begin{figure}[htbp] \centerline{\psfig{figure=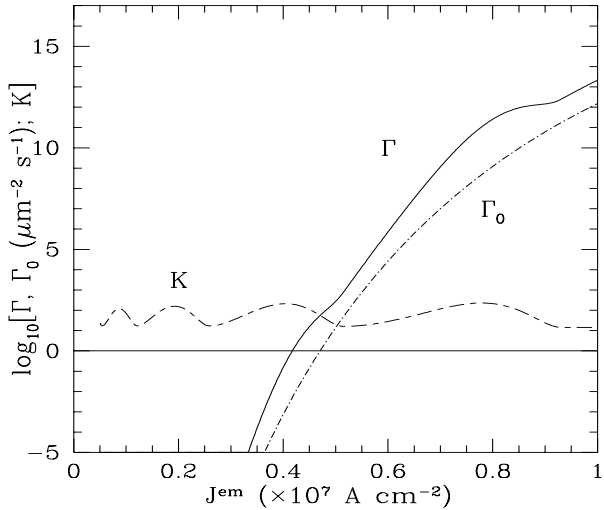,height=8cm}}
\caption{ Vortex-antivortex production rate (in $\mu m^{-2} s^{-1}$,
$\log_{10}$ scale) in the presence ($\Gamma$) and absence ($\Gamma_0$) of the
periodic pinning potential, versus current density $J^{em}$ (in
$A~cm^{-2}~\times~10^7$).  $K(J)$ is the current-dependent correction factor
due to the periodic substrate. } \end{figure}

So far, the computation referred to $T=0$. In \cite{ieju} we proposed to take
into account finite temperatures by adding incoherently, in the above formulas,
a term proportional to $T$ (note $k_B=1$):  $J^2{\rightarrow}J^2 + \frac{ \eta
{\cal E}_{0R} }{8\pi}T$, such that for $J{\rightarrow}0$ the standard Boltzmann
exponential form in the vortex density, $\rho_v{\propto}\sqrt{\Gamma}$, is
recovered in the absence of pinning. This might be a reasonable qualitative
treatment of thermal effects for temperatures less than ${\approx}8\pi J^2/\eta
{\cal E}_{0R}$. Furthermore, we propose, following Minnhagen \cite{minnh}, to
account for vortex-antivortex Coulomb-like interactions by means of a
current-dependent (and temperature- independent) activation energy: ${\cal
E}_R(J)={\cal E}_{0R} \ln (J_{max}/J)$, with ${\cal E}_{0R}$ including all
other renormalization effects. This expression entails an ${\cal E}_R$ infinite
for $J=0$ (as is appropriate below the Kosterlitz-Thouless (KT) transition in
the film) and changing sign, thus becoming unphysical, at $J_{max}$.

For illustrative purposes, in order to estimate the parameters involved in our
final expression for the rate $\Gamma$ (Eq. (\ref{final})), we take $J_{max}$
an order of magnitude higher than the single-crystal YBCO critical current,
\cite{cypa} namely $J_{max}^{em}=10^8$ A cm$^{-2}$, and take ${\cal
E}_R{\approx}80$ K for $J=10^7$ A cm$^{-2}$, of the order of magnitude of the
KT transition temperature \cite{reviews}. The important friction coefficient is
taken to be $\eta{\approx}10^{-2}$ ${\AA}^{-2}$, from the Bardeen-Stephen
formula \cite{bast} applied to YBCO films.  A most sensitive parameter is
related to the amplitude of the pinning potential, $\epsilon=A_0/{\cal
E}_{0R}^2$, which, like the pinning lattice spacing $d$, is entirely unknown.
We have taken a negative $\epsilon=-0.5$, borrowing from classical nucleation
the point of view that vortex production is actually aided by the presence of
pinning centers.  Finally, $d=50$ $\AA$ concludes our illustrative case which
is clearly of the confined-phase type. As $\eta > \eta_c$, we have taken
$\mu{\approx}1$ and $A_{0R}{\approx}A_0$, for simplicity's sake, and find no
significant dependence on the most uncertain of all parameters, $\Omega$. Its
dimensions being the same as ${\cal E}_0^2$, in Fig. 1 we have taken
$\Omega={\cal E}_{0R}^2/e$.  Also, we take the film thickness $s=10$ $\AA$ as
the typical interlayer spacing. This figure reports the production rate, both
in the absence and in the presence of the above-specified periodic pinning
potential. We stress that, despite our possibly optimal choice of parameters,
we believe that the order of magnitude of $\Gamma$ is such as to make vortex
nucleation observable within acceptable values for the current density of some
high-$T_c$ materials. When pinning is present, we predict for the first time
the possibly observable oscillations of $\Gamma (J)$, reflecting the underlying
pinning periodicity. This should produce microscopic oscillations in the
measured resistence at relatively low temperatures ($T=5$ K in Fig. 1).  For
higher temperatures, vortex nucleation would be essentially thermally activated
and therefore outside the scope of the present investigation.

We hope that these findings will stimulate further research, both theoretical
and experimental, on the quantum mechanics of vortices in thin superconducting
films.

One of us (GJ) is grateful to the International School for Advanced Studies in
Trieste, where part of this work was carried out, for hospitality and the use
of its facilities.


\begin{references}
\bibitem{note1} We assume full
isotropy in the potential, the extension to the anisotropic case being
trivial.
\end{references}
\end{document}